# Comparative Evaluation and Analysis of IAX and RSW

Manjur S Kolhar[1], Mosleh M. Abu-Alhaj[1], Omar Abouabdalla[1], Tat Chee Wan[1], and Ahmad M. Manasrah[1]
1: National Advanced IPv6 Centre of Excellence
1: Universiti Sains Malaysia
1: Penang, Malaysia
1: {manjur, mosleh, omar, tcwan, ahmad}@nav6

*Abstract*— Voice over IP (VoIP) is a technology to transport media over IP networks such as the Internet. VoIP has the capability of connecting people over packet switched networks instead of traditional circuit switched networks. Recently, the InterAsterisk Exchange Protocol (IAX) has emerged as a new VoIP which is gaining popularity among VoIP products. IAX is known for its simplicity, NAT-friendliness, efficiency, and robustness. More recently, the Real time Switching (RSW) control criterion has emerged as a multimedia conferencing protocol. In this paper, we made a comparative evaluation and analysis of IAX and RSW using Mean Opinion Score rating (MOS) and found that they both perform well under different network packet delays in ms.

## I. INTRODUCTION

Voice over IP (VoIP) is an Internet Protocol (IP) telephony that facilitates the delivery of voice packets over the Internet. VoIP sends voice information in digital form in discrete packets rather than use the traditional circuit-switched protocols of the public switched telephone network (PSTN). ITU-T recommendation H.323 [1] was the first standard for Internet telephony. Among the widely used and available VoIP protocols are Session Initialization Protocol (SIP), H.323, and IAX. The H.323 standard, approved in 1996 by the ITU Study Group 15, specifies the technical requirements for media and data communication in packet-based networks. H.323 is used for developing real-time multimedia communication, such as audio and video conferencing, over packet-switched networks. Network-based audio and video conferencing has established itself as an important element in the VoIP industry. H.323 is an umbrella specification because it includes various other ITU standards. The components under H.323 architecture are end-point terminals, gatekeeper, and multipoint control unit (MCU). The Session Initiation Protocol (SIP) is a signaling protocol for initiating, managing, and terminating voice and video sessions across packet networks. SIP is being developed by the SIP Working Group within the Internet Engineering Task Force (IETF). SIP sessions involve one or more participants and can use unicast or multicast communication. SIP is a protocol that inherits a lot of its designs from HTTP and SMTP. SIP can be extended to accommodate extra features and services such as call control services, mobility, and interoperability with existing telephony systems. The protocol is published as IETF RFC 2543 and currently has the status of a proposed standard. SIP is growing in popularity due to its capability to easily combine voice and IP-based services. IAX is an interesting alternative besides communication protocols, which is used nowadays by service providers in their conversational service offerings (e.g., H.323 and SIP). It is used for both signaling and media-control operations. Moreover, it provides interesting features such as management of signaling and media transfer. IAX is a simple protocol which has the capability of avoiding NAT traversal complications. Further, there are many features [2] that the IAX protocol offers, which are unavailable in other existent VoIP signaling protocols.

The RSW Control Criteria is based on bandwidth reduction techniques [3]. The RSW Multimedia Conference is comprised of the conference chairman, participants, and passive observers. The chairman of the conference is the organizer of the conference, while the other conference members can be participants, passive observers, or simply observers. The objective of this article is to experimentally study and analyze IAX protocol performance, and compare it with RSW. An experimental study on live VoIP traffic was carried out using Asterisk open-source PBX, and for RSW, a Multimedia Conferencing System called MCS was employed [4].

## II. COMPARATIVE STUDY: IAX AND RSW

We already proposed the concept of interworking between these two protocols to maximize return on current investments [5]. We compare these to services in terms of signaling and media translation. In the following sub-sections, we analyze their differences, which are believed to be more important.

*A. Signaling*

The signaling session is the first to be performed by any VoIP protocol before transmission of any media types, and it is one of the most important functions in the VoIP infrastructure. This is because it allows various network components to communicate with one another, to set up, and to tear down calls.

As shown in Table 1, to initiate a conference or to invite for a conference in RSW, the CREATE signal is used. CREATE can also be used describe the media types to be used during a conference. The RSW server extracts the CREATE signal and sends the invitation call to the destination endpoint. The destination replies with JOIN, REJECT, or BUSY. The participant can leave the conference by sending LEAVE. Only the Chairman is allowed to END the conference.

TABLE I. FONT SIZES FOR PAPERS

| Signals | Usage |
|---|---|
| CREATE | To create session |
| JOIN | To join session |
| LEAVE | To leave session |
| END | To end session |
| ACK | Acks' |

Table 2 shows the IAX protocol sessions between the IAX endpoints. The initial singling process starts with a NEW message indicating the destination number. The remote peer can respond with either a credentials challenge (AUTHREQ), a REJECT message, or an ACCEPT message. The AUTHREQ message indicates the permitted authentication schemes and SHOULD result in the sending of an AUTHREP message with the requested credentials. The REJECT message indicates that the call cannot be established at this time. ACCEPT indicates that the call leg between these two peers is established. Typically, the first call control message is RINGING, but a PROCEEDING message MAY precede it or the call MAY proceed directly to the ANSWER message.

The RSW and IAX signaling techniques appear to be relatively similar in nature. The messages for both protocols can be grouped into two categories: "requests" and "responses" [5].

### B. Media

RSW and IAX have very different approaches for supporting the exchange of media packets. In the case of real-time transmission of voice or video, the commonly used media transfer protocol is RTP. Once the singling session is established, RSW makes RTP take the responsibility of transferring media data. Meanwhile, IAX uses UDP for both signaling and data, and it employs mini frame architecture to send media data between the ongoing media conference. These mini-frames only have a 4-byte header which is composed of the source call number and the lower two octets of the timestamp.

TABLE II. IAX SIGNALING MESSAGES

| Signals | Usage |
|---|---|
| NEW | To place call |
| AUTHREQ | To authenticate |
| ACCEPT | To accept call leg |
| PROCEEDING | Proceed to join |
| RINGING | Ring at destination |
| Answer | In Call |

## III. EXPERIMENTS

We used Asterisk for our experimental study, open source software based on PBX, which supports all VoIP protocols along with many voice codec.

### A. Method for Quality Assessment

IP-based communication technologies are progressing rapidly, so it is very important to assess them in terms of QoS for diagnostic system design, and effective network layouts [6].

To measure network performance, we used the Mean Opinion Score (MOS) rating, which is the most widely used assessment technique. It is standardized by ITU-T in Recommendation P.800 [9]. The listening subjects were given a scale from 1 to 5, where 1 = bad, 2 = poor, 3 = fair, 4 = good, and 5 = excellent [7].

### B. Experimental Setup

Figure 1 shows the experimental set-up used to conduct the study. Both the servers RSW and Asterisk are running on a single Linux machine. To induce delay in the used network, we used netem, a network emulation tool for testing protocol behavior by emulating the property of a wide area network. Netem emulates the variable delay, duplication, and reordering of network packets, but we e used network delay option only.

Our experiments focused on comparing the performance of RSW and IAX in the presence of packet delay. To examine quality effects due to packet delay, both protocols were compared with a fixed packet delay ranging from 0 to 2000 ms with increments of 25 ms. As Figure 2 indicates, the IAX protocol performs slightly better in the presence of fixed packet delay.

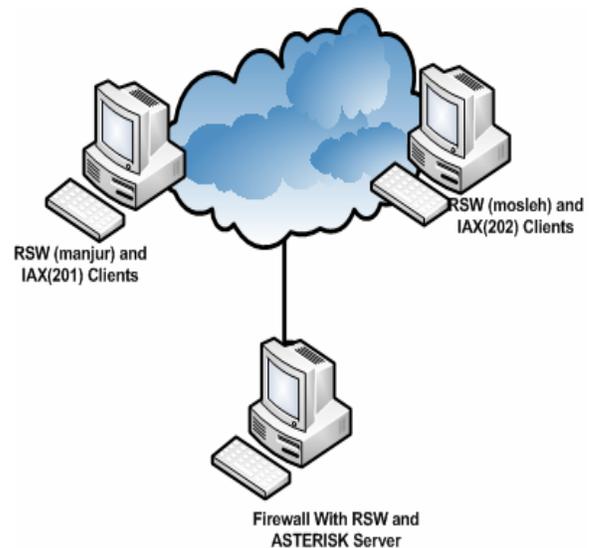

Figure 1. Experimental Set-up

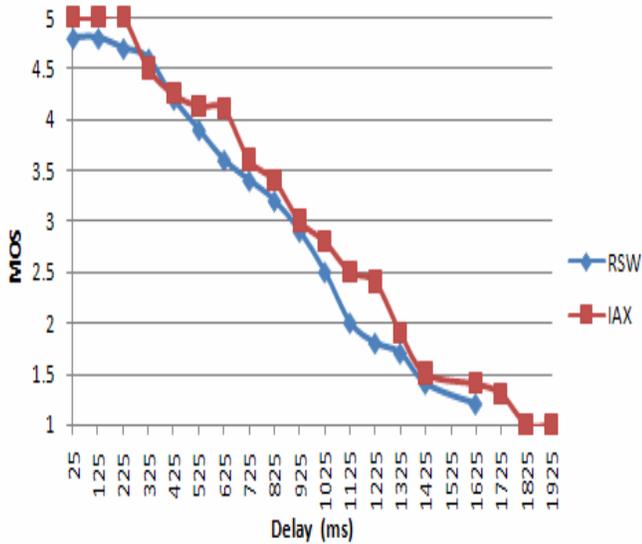

Figure 2. Packet Delay – RSW versus IAX

## CONCLUSIONS

To the best of our knowledge, our work is one of the first to examine RSW and IAX in the presence of packet delay. We presented comparative studies on IAX and RSW in the presence of packet delay in the network. Our experimental results showed that both IAX and RSW perform better in network delays. This is due to IAX's mini frame architecture which gave a slight edge to IAX when the network delay is in the range of 1425 to 1925. In order to cross-check our experimental setup and results, further research on IAX and RSW control criteria protocols and their performance under various conditions is recommended.

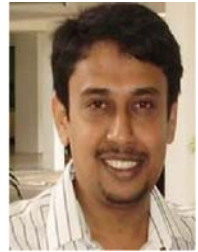

**Mr. Manjur S Kolhar,** did his Masters in Computer Application, and thereafter worked with National Aerospace Limited, Bangalore India. In 2002 Joined MISTRAL, Bangalore, INDIA. Later he Joined U&I System Design as a Member technical Software Developer. Now working with MLABS SDN BHD Penang, Malaysia on project bases and work involves on multimedia streaming, Microsoft based direct show, direct sound and on hardware based High Definition Video streaming.

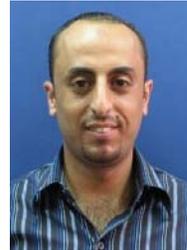

**Mosleh M. Abu-Alhaj** is a researcher pursuing his PhD in Computer Science at the National Advanced IPv6 Center of Excellence in University Sains Malaysia. He received his first degree in Computer Science from Philadelphia University, Jordan, in July 2004 and his Master degree in Computer Information System from the Arab Academy for Banking and Financial Sciences, Jordan in July 2007. His research area of interest includes VoIP, Multimedia Networking, and Session initiation Protocol. Apart from research, he also does consultancy services in the above research areas and directs the VoIP team at NAv6.

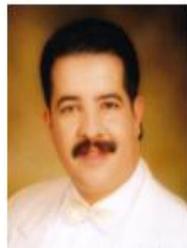

**Dr. Omar Amer Abouabdalla** is a senior lecturer and head of the technical department in the National Advanced IPv6 Centre (NAv6) - University Science Malaysia (USM). Dr. Omar is the Chairman of multimedia working group (a sub working groupin APAN), Asia Pacfic Advanced Network (APAN) is a high bandwidth network that will interconnect the Asia Pacifc Countries. He is also a member of Internet Engineering Task Force (IETF) and a member of Editorial Board for Journal of IT in Asia. Dr. Omar is heavily involved in researches carried by NAv6 center, such as Multimedia Conferencing System (MCS) and IPv6 over Fiber project. He has more than five years experience in the field of IPv6 and more than ten years in the field of Multimedia Network.

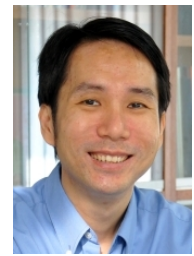

**Tat-Chee Wan** received his BSEE and MSECE from University of Miami, FL in 1990 and 993 respectively, and his Ph.D from Universiti Sains Malaysia, Penang, Malaysia, in 2005. He is currently the Programme Chairman for Computer Systems at the School of Computer Sciences, Universiti Sains Malaysia, Penang, Malaysia. His research interests include Satellite, Wireless and Sensor Networks, Multicast protocols, QoS and embedded real-time systems.

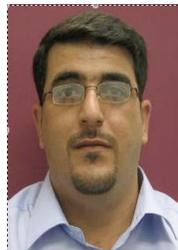

**Dr. Ahmed M. Manasrah** is a senior lecturer and the Head of iNetmon project as well as the research and innovation of the National Advanced IPv6 Centre of Excellence (NAV6) in Universiti Sains Malaysia. He is also the IMPACT Research Domain Head for Botnet and threat assessment Research. Dr. Ahmed obtained his Bachelor of Computer Science from Mu'tah University, al Karak, Jordan in 2002. He obtained his Master of Computer Science and doctorate from Universiti Sains Malaysia in 2005 and 2009 respectively. Dr. Ahmed is heavily involved in researches carried by NAv6 centre, such as Network monitoring and Network Security monitoring with filling 3 Patents in Malaysia.